\documentclass[12pt]{article}
\usepackage{amsfonts}
\usepackage{amsmath}
\usepackage{hyperref}

\newcommand{\ga}{\gamma}
\newcommand{\om}{\omega}
\newcommand{\ep}{\varepsilon}
\newcommand{\dl}{\delta}
\newcommand{\rmd}{{\rm d}}
\newcommand{\rme}{{\rm e}}
\newcommand{\rmi}{{\rm i}}
\newcommand{\Tr}{{\rm Tr}\,}
\newcommand{\refcite}{\cite}

\newtheorem{theorem}{Theorem}
\newtheorem{definition}{Def.}
\newtheorem{remark}{Remark}

\begin{document}

\title{White noise approach to the low density limit\\of a quantum particle in a gas}
\addcontentsline{toc}{section}{Title Page}

\author{Alexander Pechen\thanks{This preprint is a minor modification of the
paper published in {\it QP-PQ: Quantum Probability and
White Noise Analysis} {\bf 18}, Eds. M. Sch\"urmann \& U.
Franz, (2005) 428--447. The Appendix with several comments
is added. The paper was written while the author visited
the Centro Vito Volterra of Rome University Tor Vergata,
Rome, Italy in February-March 2004 on leave from Steklov
Mathematical Institute of Russian Academy of
Sciences.}\\
Department of Chemistry, Princeton University, Princeton,
NJ 08544\\ E-mail: apechen@princeton.edu}

\maketitle

\abstract{The white noise approach to the investigation of
the dynamics of a quantum particle interacting with a
dilute and in general non-equilibrium gaseous environment
in the low density limit is outlined. The low density
limit is the kinetic Markovian regime when only pair
collisions (i.e., collisions of the test particle with one
particle of the gas at one time moment) contribute to the
dynamics. In the white noise approach one first proves
that the appropriate operators describing the gas converge
in the sense of appropriate matrix elements to certain
operators of quantum white noise. Then these white noise
operators are used to derive quantum white noise and
quantum stochastic equations describing the approximate
dynamics of the total system consisting of the particle
and the gas. The derivation is given {\it ab initio},
starting from the exact microscopic quantum dynamics. The
limiting dynamics is described by a quantum stochastic
equation driven by a quantum Poisson process. This
equation then applied to the derivation of quantum
Langevin equation and linear Boltzmann equation for the
reduced density matrix of the test particle. The first
part of the paper describes the approach which was
developed by L. Accardi, I.V. Volovich and the author and
uses the Fock-antiFock (or GNS) representation for the CCR
algebra of the gas. The second part presents the approach
to the derivation of the limiting equations directly in
terms of the correlation functions, without use of the
Fock-antiFock representation. This approach simplifies the
derivation and allows to express the strength of the
quantum number process directly in terms of the
one-particle $S$-matrix.}

\section{Introduction}
The fundamental equations in quantum theory are the
Heisenberg and Schr\" odinger equations. However, it is a
very difficult problem to solve explicitly these equations
for realistic physical models and one uses various
approximations or limiting procedures such as weak
coupling and low density limits. These scaling limits
describe the long time behavior of physical systems in
different physical regimes.

One of the methods to study the long time behavior in
quantum theory is the stochastic limit method, which was
developed by Accardi, Lu and Volovich in the
book~\cite{ALV}. The book is devoted mainly to the weak
coupling regime, where one considers the long time
dynamics of a quantum open system weakly interacting with
a reservoir. The dynamics of the total system in this
limit is described by the solution of a quantum stochastic
differential equation driven by a quantum Brownian motion;
the reduced dynamics of the system is described by a
quantum Markovian master equation. Also in the low density
regime the dynamics is described by a quantum stochastic
differential equation. In this regime one considers the
long time quantum dynamics of a test particle interacting
with a dilute gas (we consider Bose gas in this paper) in
the case the interaction is not weak but the density $n$
of particles of the gas is small. The {\it low density
limit} is the limit as $n\to 0$, $t\to\infty$, $nt=Const$
($t$ denotes time) of certain quantities such as matrix
elements of the evolution operator or the reduced density
matrix. In this limit the reduced time evolution for the
test particle will be Markovian, since the characteristic
time $t_S$ for appreciable action of the gas on the test
particle (time between collisions) is much larger than the
characteristic time $t_R$ for relaxation of correlations
in the gas.

Accardi and Lu~\cite{AcLu,AL2,L} and later Rudnicki,
Alicki, and Sadowski~\cite{ras} proved that the matrix
elements of the evolution operator in the "collective
vectors" converge in the low density limit to matrix
elements of the solution of a quantum stochastic
differential equation driven by a quantum Poisson process.
The quantum Poisson process, introduced by Hudson and
Parthasarathy~\cite{HP}, should arise naturally in the low
density limit, as conjectured by Frigerio and
Maassen~\cite{FrMa}.

The stochastic golden rule for the low density limit, that
is a set of simple rules for the derivation of the
limiting quantum white noise and stochastic differential
equations, was developed in Refs~\cite{APV1,APV2,p}, where
the case of a discrete spectrum of the free test
particle's Hamiltonian was considered. This method is
based on the white noise approach and uses the stochastic
limit technique~\cite{ALV} (in Ref.~\cite{lp} the method
is generalized to the case of a continuous spectrum). As
the main result, the normally ordered quantum white noise
and stochastic differential equations were derived. Then
these equations were applied to the derivation of the
quantum Langevin and master equations describing the
evolution of the test particle.

The idea of the white noise approach is based on the fact
that the rescaled free evolution of appropriate operators
of the gas converges, in the sense of convergence of
correlation functions, to certain operators of quantum
white noise. Then, using these limiting operators which
are called master fields, one can derive the quantum white
noise equation for the limiting dynamics and put this
equation to the normally ordered form which is equivalent
to a quantum stochastic differential equation.

The white noise approach to the derivation of the
stochastic equations in
Refs.~\refcite{AcLu,AL2,L,APV1,APV2} uses the
Fock-antiFock representation for the canonical commutation
relations (CCR) algebra of the Bose gas, which is unitary
equivalent to Gel'fand-Naimark-Segal (GNS) representation.
The white noise approach of Ref.~\refcite{p} considerably
simplifies the calculations by avoiding use of the
Fock-antiFock representation. A useful tool is the energy
representation introduced in Refs.~\refcite{APV1,APV2},
where the case of orthogonal formfactors was considered.
This consideration was extended in Ref.~\refcite{p} to the
case of arbitrary formfactors and arbitrary, not
necessarily equilibrium, quasifree low density states of
the gas.

In Ref.~\refcite{p} to each initial low density state of
the Bose gas in the low density limit one associates a
special "state", which is called a causal state, on the
limiting master field algebra. The time-ordered (or
causal) correlators of the initial Bose field converge to
the causal correlators of the master fields, which are
number operators of quantum white noise. This approach
allows to express the intensity of the quantum Poisson
process directly in terms of the one-particle $S$-matrix.
In this case the algebra of the master
fields~(\ref{ccrB0}), the limiting
equation~(\ref{normordeq1}), and the quantum Ito
table~(\ref{ito}) do not depend on the initial state of
the Bose gas (see Sect.~\ref{sect4} for details). Instead,
the information on the initial state of the gas is
contained in the limiting state $\varphi_L$ of the master
field [defined by~(\ref{prop1})--(\ref{state1})], which is
now not the vacuum. Here the operator $L$ determines the
initial density of particles of the gas; if the gas is at
equilibrium with inverse temperature $\beta$, then
$L=e^{-\beta H_1}$ (see the next section). To get the
master equation one has to take the conditional
expectation with respect to the state $\varphi_L$.

The dynamics in the low density limit is given by the
solution of the quantum white noise
equation~(\ref{normordeq1}) or of the equivalent quantum
stochastic equation
\begin{equation}\label{1}
\rmd U_t=\rmd N_t(S-1)U_t
\end{equation}
where $U_t$ is the evolution operator at time $t$, $S$ the
one-particle $S$ matrix describing scattering of the test
particle on one particle of the gas, and $N_t(S-1)$ the
quantum number process with strength $S-1$. In order to
describe these objects let us introduce two Hilbert spaces
${\mathcal H}_{\rm S}$ and ${\mathcal H}_1$, which are
called in this context the system and one-particle
reservoir Hilbert spaces, and the Fock space
$\Gamma(L^2(\mathbb R_+;{\mathcal H}_1))$ over the Hilbert
space of square-integrable measurable vector-valued
functions from $\mathbb R_+=[0,\infty)$ to ${\mathcal
H}_1$. Then the solution of the equation is a family of
operators $U_t; t\ge 0$ in ${\mathcal H}_{\rm
S}\otimes\Gamma(L^2(\mathbb R_+;{\mathcal H}_1))$ (adapted
process); $S$ is a unitary operator in ${\mathcal H}_{\rm
S}\otimes{\mathcal H}_1$.

Let us define the number process. Let $X$ be a
self-adjoint operator in a Hilbert space $\mathcal K$; for
any $f\in{\mathcal K}$ let $\Psi(f)$ be the normalized
coherent vector in the Fock space $\Gamma({\mathcal K})$.
The {\it number operator} $N(X)$ is the generator of the
one-parameter unitary group $\Gamma(e^{itX})$
characterized by $\Gamma(e^{itX})\Psi(f)=\Psi(e^{itX}f)$;
$t\in\mathbb R$. The number operator is characterized by
the property $\langle\Psi(f),N(X)\Psi(g)\rangle=\langle
f,Xg\rangle\langle\Psi(f),\Psi(g)\rangle$. The definition
of $N(X)$ is extended by complex linearity to any bounded
operator $X$ on $\mathcal K$. Let us consider $\mathcal K$
of the form $L^2(\mathbb R_+;{\mathcal H}_1)\cong
L^2(\mathbb R_+)\otimes{\mathcal H}_1$. For any bounded
operators $X_0\in B({\mathcal H}_{\rm S})$, $X_1\in
B({\mathcal H}_1)$ and for any $t\ge 0$ define
$N_t(X_0\otimes X_1):=X_0\otimes N(\chi_{[0,t]}\otimes
X_1)$ and extend this definition by linearity to any
bounded operator $K$ in ${\mathcal H}_{\rm
S}\otimes{\mathcal H}_1$. The family $\{N_t(K)\}_{t\ge 0}$
of operators in ${\mathcal H}_{\rm
S}\otimes\Gamma(L^2(\mathbb R_+;{\mathcal H}_1))$ is
called {\it quantum number process with strength $K$}.

Equations~(\ref{1}) and~(\ref{normordeq}) describe the
dynamics of the total system and can be applied, in
particular, to the derivation of the irreversible quantum
linear Boltzmann equation for the reduced density matrix
of the test particle. This equation can be directly
obtained from the quantum Langevin equation~\cite{APV2}.

The reduced dynamics of the test particle in the low
density limit with methods, based on a quantum
Bogoliubov-Born-Green-Kirkwood-Yvon (BBGKY) hierarchy, has
been investigated by D\"umcke~\cite{dumcke}, where it is
proved that, under some conditions, the reduced dynamics
is given by a quantum Markovian semigroup.

In the white noise approach the reduced dynamics can be
directly derived from the solution of the limiting quantum
stochastic differential equation. Namely, the limiting
evolution operator $U_t$ and the limiting state
$\varphi_L$ determine the reduced dynamics by
\begin{equation}\label{T1}
T_t(X)=\varphi_L(U^+_t(X\otimes 1)U_t),
\end{equation}
where $X$ is any observable of the test particle,
$\varphi_L(\cdot)$ denotes the conditional expectation,
and $T_t$ is the limiting semigroup. This equality shows
that $U_t$ is a stochastic dilation of the limiting
Markovian semigroup. Using the quantum Ito table for
stochastic differential ${\rm d}N_t$ one can derive a
quantum Langevin equation for the quantity $U^+_t(X\otimes
1)U_t$. Then, taking partial expectation, one gets an
equation for $T_t(X)$ and obtains the generator of the
semigroup (see the end of Sec.~\ref{sect4}). This is a
general rule of the white noise approach: one first
obtains the Langevin equation and then gets the reduced
dynamics of the test particle. Let us note that although
the quantum stochastic equation~(\ref{normordeq}), which
was derived in Refs.~\refcite{AcLu,APV2}, is different
from~(\ref{1}) it gives the same reduced dynamics.

The low density limit can be applied to the model of a
test particle moving through an environment of randomly
placed, infinitely heavy scatterers (Lorentz gas) (see the
review of Spohn\cite{spohn}). In the Boltzmann--Grad limit
successive collisions become independent and the averaged
over the positions of the scatterers the position and
velocity distribution of the particle converges to the
solution of the linear Boltzmann equation. An advantage of
the stochastic limit method is that it allows us to derive
equations not only for averaged over reservoir degrees of
freedom dynamics of the test particle but for the total
system+reservoir. The convergence results and derivation
of the linear Boltzmann equation for a quantum Lorentz gas
in the low density and weak coupling limits are presented
in Refs.~\refcite{EPT,EY1}. The Coulomb gas at low density
is considered in Ref.~\refcite{CLY}.

The main results presented in this paper are: the causally
normally ordered quantum white noise
equations~(\ref{normordeq}), (\ref{equ1}) and equivalent
quantum stochastic equation~(\ref{qsde3}) for the limiting
evolution operator; the quantum Langevin
equation~(\ref{Langevin}) for the evolution of any test
particle's observable; the linear Boltzmann equation for
the reduced density matrix (Theorem~\ref{thb}).

The structure of the paper is the following. In Sec.~2 a
test particle interacting with a Bose gas is considered.
In Sec.~3 the white noise approach developed by L.
Accardi, I. Volovich, and the author in
Refs.~\refcite{APV1,APV2} is presented. Sec.~4 describes
the white noise approach developed in Ref.~\refcite{p}.

\section{Test Particle Interacting with a Dilute Bose Gas}
Consider two non-relativistic particles, with masses $M$
and $m$, which are called the test particle and a particle
of the gas. Suppose the particles interact by a pair
potential $U(R-r)$, where $R$ and $r$ denote positions of
particles. Then the classical dynamics of the particles is
determined by the Hamiltonian $H_{\rm
cl}=P^2/2M\,+p^2/2m\,+ U(R-r)$, where $P$ and $p$ are
momentums of the particles.

The quantum Hamiltonian of such system is obtained by
identification of $P$ with the momentum operator $\hat P$,
$R$ with the position operator $\hat Q$ and, if instead of
one particle of mass $m$ there is gas of these particles,
by second quantization of the particles of the gas. This
Hamiltonian has the form $H=H_S+H_R+H_{\rm int}$, where
$H_S+H_R=:H_0$ is the free Hamiltonian with $H_S=\hat
P^2/2M$, $H_R=\int\omega(p)a^+(p)a(p)dp$,
$\omega(p)=p^2/2m$; the interaction Hamiltonian is $
H_{\rm int}=\int U(\hat Q-r)a^+(r)a(r)dr$. The free
Hamiltonian of one particle of the gas $H_1$ is the
multiplication operator by the function $\om$. Boson
annihilation and creation operators $a(k),a^+(p)$ satisfy
the canonical commutation relations
$$
[a(k),a^+(p)]=\delta(k-p)
$$
The coordinate representation for these operators is
introduced as $ a(r)=\int e^{ikr}a(k)dk$. The interaction
Hamiltonian can be written using the Fourier transform of
the interaction potential $ \tilde U(p):=\int
U(r)e^{ipr}dr$ as
\begin{equation}\label{Hint0}
H_{\rm int}=\int dk dp\tilde U(p)e^{ip\hat Q}\otimes a^+(k-p)a(k)
\end{equation}
This Hamiltonian acts in the Hilbert space ${\mathcal
H}_{\rm S}\otimes\Gamma({\mathcal H}_1)$, where ${\mathcal
H}_{\rm S}={\mathcal H}_1=L^2(\mathbb R^3)$.

Important features of this system are that the interaction
Hamiltonian $H_{\rm int}$ quadratic in creation and
annihilation operators and commutes with the number
operator, i.e., it preserves the number of particles of
the gas.

We will consider, instead of~(\ref{Hint0}), a different
interaction Hamiltonian, which nevertheless keeps its
basic properties. More precisely, we consider Hamiltonians
of the form
\[
H_{\rm int}:=D\otimes A^+(g_0)A(g_1)+D^+\otimes
A^+(g_1)A(g_0)
\]
where $D$ is a bounded operator in ${\mathcal H}_{\rm S}$; $g_0, g_1\in{\mathcal H}_1$ are two form-factors, and
$A(g_0)=\int dk g_0^*(k)a(k)$ is the smeared annihilation operator. This Hamiltonian is also quadratic in creation
and annihilation operators and preserves the number of particles of the gas. In the present paper we consider the
case of a discrete spectrum for the free test particle's Hamiltonian, so that
\begin{equation}\label{1.1}
H_{\rm S}=\sum\limits_n\ep_nP_n
\end{equation}
where $\ep_n$ is an eigenvalue and $P_n$ is the corresponding projector. This corresponds to the situation when
the test particle is confined in some spatial region.

Density of particles of the gas is encoded in the state of
the gas, which is chosen to be either the Gibbs state at
inverse temperature $\beta$, chemical potential $\mu$, and
fugacity $\xi=e^{\beta\mu}$, or a more general
non-equilibrium Gaussian state, i.e., a gauge invariant
mean zero Gaussian state with the two point correlation
function
\begin{equation}\label{state}
\varphi_{L,\xi}(A^+(f)A(g))=\xi\left\langle g,\frac{L}{1-\xi L}f\right\rangle
\end{equation}
Here $\xi>0$ is a small positive number and $L$ is a
bounded positive operator in ${\mathcal H}_1$ commuting
with the one-particle free evolution $S_t=e^{itH_1}$ [the
multiplication operator by a function $L(k)$]. In the case
$L=\rme^{-\beta H_1}$, so that
$L(k)=\rme^{-\beta\omega(k)}$, the state $\varphi_{L,\xi}$
is just the Gibbs state with the two point correlation
function
\[
\varphi_{L,\xi}(a^+(k)a(k'))=n(k)\dl(k-k')
\]
where $n(k)$ is the density of particles of the gas with momentum $k$:
\[
n(k)=\frac{\xi e^{-\beta k^2/2m}}{1-\xi e^{-\beta k^2/2m}}
\]
Notice that in the limit $\xi\to 0$ the density goes to zero. Therefore the limit $\xi\to 0$ is equivalent to the
limit $n(k)\to 0$.

The dynamics of the total system is determined by the evolution operator which in interaction representation has
the form $U(t):=\rme^{\rmi tH_{\rm free}}\rme^{-\rmi tH_{\rm tot}}$. The evolution operator satisfies the
differential equation
\[
 \frac{dU(t)}{dt}=-\rmi H_{\rm int}(t)U(t),
\]
where $ H_{\rm int}(t)=\rme^{\rmi tH_{\rm free}} H_{\rm int}\rme^{-\rmi tH_{\rm free}}$ is the free evolution of
the interaction Hamiltonian. The iterated series for the evolution operator is
\begin{equation}\label{eqU111}
U(t)=1+\sum\limits_{n=1}^\infty(-\rmi)^n\int\limits_0^tdt_1\dots\int\limits_0^{t_{n-1}}dt_n
H_{\rm int}(t_1)\dots H_{\rm int}(t_n)
\end{equation}
With the notation $ D(t):=\rme^{\rmi tH_{\rm
S}}D\rme^{-\rmi tH_{\rm S}}$ the evolved interaction
becomes
\[
H_{\rm int}(t):= D(t)\otimes
A^+(S_tg_0)A(S_tg_1)+D^+(t)\otimes A^+(S_tg_1)A(S_tg_0).
\]
Using the spectral decomposition~(\ref{1.1}) and introducing the set of all Bohr frequencies $B$, that is the
spectrum of the free test particle's Liouvillean $i[H_{\rm S},\cdot]$, one can write the free evolution of $D$ as
\[
D(t) =\sum\limits_{\om\in B} D_\om e^{-it\om};\qquad D_\om=\sum\limits_{k,m: \ep_m-\ep_k=\omega}P_kDP_m
\]

The reduced dynamics of any test particle's observable $X$ in the low density limit is defined as the limit
\[
T_t(X):=\lim\limits_{\xi\to 0}\varphi_{L,\xi}(U^+(t/\xi)(X\otimes 1)U(t/\xi))
\]
where $\varphi_{L,\xi}(\cdot)$ denotes partial expectation. The reduced density matrix $\rho(t)$ is defined
through the duality $\Tr(\rho(0)T_t(X))=\Tr(\rho(t)X)$. As it was mentioned in the Introduction, in the white
noise approach the generator of the limiting semigroup can be easily derived from the quantum white noise
equation.

\section{The White Noise Approach}\label{sect3}
In Refs.~\refcite{AcLu,AL2,APV2} the dynamics of the total system is constructed in the Fock-antiFock
representation for the CCR algebra, which is unitary equivalent to the GNS representation. It is defined as
follows.

Denote by ${\mathcal H}_1^\iota$ the conjugate of ${\mathcal H}_1$, i.e. ${\mathcal H}_1^\iota$ is identified to
${\mathcal H}_1$ as a set and the identity operator $\iota :\ {\mathcal H}_1\longrightarrow {\mathcal H}_1$ is
antilinear: $\forall f\in{\mathcal H}_1,\, c\in\mathbb C$
$$
\qquad \iota(c f)=c^* \iota (f),\qquad \langle\iota(f),\iota(g)\rangle_\iota =\langle g,f\rangle
$$
Then, ${\mathcal H}_1^\iota $ is a Hilbert space and, if the vectors of ${\mathcal H}_1$ are thought as
ket-vectors $|\xi\rangle$, then the vectors of ${\mathcal H}_1^\iota$ can be thought as bra-vectors $\langle\xi|$.
The corresponding Fock space $ \Gamma({\mathcal H}_1^\iota) $ is called the anti-Fock space.

In this section we assume that for any $t\in\mathbb R$: $\langle g_0,S_tg_1\rangle=0$. In this case it was shown
in Ref.~\refcite{AcLu} that the dynamics of the total system is given by the family of unitary operators
$U_t^{(\xi)}$ in $ {\mathcal H}_S\otimes\Gamma({\mathcal H}_1)\otimes \Gamma({\mathcal H}_1^\iota) $ which satisfy
the Schr\"odinger equation:
$$
\partial_tU_t^{(\xi)}=-iH_\xi(t)U_t^{(\xi)},\qquad U_0^{(\xi)}=1.
$$
Here the part of the modified Hamiltonian, which gives a nontrivial contribution in the low density limit, has the
form
\begin{eqnarray*}
H_\xi(t)&=&\sum\limits_{\omega\in B} D_{\omega}e^{-it\omega}\otimes\Bigl\{A^+(S_tg_0)A(S_tg_{1})\otimes 1\\
&&+\sqrt{\xi}\bigl[A(S_tg_{1})\otimes A(S_te^{-\beta
H_1/2}g_0)+A^+(S_tg_0)\otimes A^+(S_te^{-\beta
H_1/2}g_{1})\bigr]\Bigr\}+h.c.
\end{eqnarray*}

After the time rescaling $t\to t/\xi$ the evolution operator satisfies the equation
$$
\frac{dU_{t/\xi}^{(\xi)}}{dt}=-i\sum\limits_{\omega\in B}\Bigl\{D_\omega
 \otimes\bigl[N_{0,1,\xi}(\omega,t)
 +B_{1,0,\xi}(-\omega,t)
 +B^+_{0,1,\xi}(\omega,t)\bigr]+h.c.\Bigr\}U^{(\xi)}_{t/\xi }
$$
where we introduce for each $n,m=0,1$ and $\om\in B$ the rescaled fields:
\begin{eqnarray}
 N_{n,m,\xi}(\omega,t)&:=&\frac{1}{\xi }e^{-it\omega/\xi }A^+(S_{t/\xi }g_{n})
 A(S_{t/\xi }g_{m})\otimes 1\label{dfn}\\
 B_{n,m,\xi}(\omega,t)&:=&\frac{1}{\sqrt{\xi}}e^{it\omega/\xi }
 A(S_{t/\xi }g_{n})\otimes A(S_{t/\xi }e^{-\beta H_1/2}g_{m})\label{dfb+}
\end{eqnarray}
and $B^+_{n,m,\xi}(\om,t)$ is the adjoint of $B_{n,m,\xi}(\om,t)$.

\subsection{Master Field}\label{ssMF}
It is convenient to use the energy representation for the investigation of the limit as $\xi\to 0$ of the rescaled
fields~(\ref{dfn}), (\ref{dfb+}). It is defined in terms of the projections
\[
P_E:=\dl(H_1-E)
\]
which satisfy the properties
\[
P_EP_{E'}=\dl(E-E')P_E,\qquad P^*_E=P_E,\qquad S_t=\int
dEP_Ee^{itE}
\]
Define the energy representation for the fields~(\ref{dfn}), (\ref{dfb+}) as
\begin{eqnarray}
N_{n,m,\xi}(E_1,E_2,\omega,t)&:=&\frac{e^{it(E_1-E_2-\omega)/\xi }}{\xi } A^+(P_{E_1}g_{n})A(P_{E_2}g_{m})\otimes
1\label{N}\\
B_{n,m,\xi}(E_1,E_2,\om,t)&:=& \frac{e^{it(E_2-E_1+\omega)/\xi }}{\sqrt{\xi}}A(P_{E_1}g_{n})\otimes
A(P_{E_2}e^{-\beta H_1/2}g_{m})\hphantom{drr}\label{B}
\end{eqnarray}
and let $B^+_{n,m,\xi}(E_1,E_2,\om,t)$ be the adjoint of $B_{n,m,\xi}(E_1,E_2,\om,t)$. The operators
(\ref{dfn}),(\ref{dfb+}) can be expressed in terms of ~(\ref{N}),(\ref{B}) as
\begin{eqnarray*}
N_{n,m,\xi}(\omega,t)&=&\int dE_1dE_2N_{n,m,\xi}(E_1,E_2,\omega,t)\\
B_{n,m,\xi}(\omega,t)&=&\int dE_1dE_2B_{n,m,\xi}(E_1,E_2,\omega,t)
\end{eqnarray*}

Let $\Omega$ be the set of all linear combinations of the Bohr frequencies with integer coefficients:
$\Omega=\{\omega\,|\,\omega=\sum_kn_k\omega_k\ {\rm with}\, n_k\in\mathbb Z,\, \omega_k\in{\rm B}\}$. Extend the
definition of the fields~(\ref{N}),(\ref{B}) to arbitrary $\om\in\Omega$. The limit as $\xi\to 0$ of the
fields~(\ref{N}),(\ref{B}) was found in Ref.~\refcite{APV2} and is given by the following theorem.
\begin{theorem}\label{mastfi}
The limits of the rescaled fields
$$
 X_{n,m}(E_1,E_2,\omega,t):=
 \lim\limits_{\xi\to 0}X_{n,m,\xi}(E_1,E_2,\omega,t), \qquad X=B,B^+,N
$$
exist in the sense of convergence of correlators and satisfy the commutation relations
\begin{eqnarray}
&[B_{n,m}(E_1,E_2,\omega,t),B^+_{n',m'}(E_3,E_4,\omega',t')]=
2\pi\dl_{\omega,\omega'}\dl_{n,n'}\dl_{m,m'}\dl(t'-t)& \nonumber\\
&\times\dl(E_1-E_3)\dl(E_2-E_4) \dl(E_1-E_2-\omega) \langle g_{n},P_{E_1}g_{n}\rangle\langle
g_{m},P_{E_2}e^{-\beta H_1}g_{m}\rangle\hphantom{ddd} & \label{crmf1}\\
& [B_{n,m}(E_1,E_2,\omega,t),N_{n',m'}(E_3,E_4,\omega',t')]= 2\pi\dl_{n,n'}\dl(t'-t)& \nonumber\\
& \times\dl(E_1-E_3)\dl(E_1-E_2-\omega) \langle g_{n},P_{E_1}g_{n}\rangle
 B_{m',m}(E_4,E_2,\omega-\omega',t)& \label{crmf2}\\
& [N_{n,m}(E_1,E_2,\omega,t),N_{n',m'}(E_3,E_4,\omega',t')]= 2\pi\dl(t'-t)& \nonumber\\
& \times\{\dl_{m,n'}\dl(E_2-E_3)\dl(E_3-E_1+\omega)\langle g_{m},P_{E_2}g_{m}\rangle
N_{n,m'}(E_1,E_4,\omega+\omega',t)& \nonumber\\
& -\dl_{n,m'}\dl(E_1-E_4)\dl(E_3-E_1-\omega')\langle g_{n},P_{E_1}g_{n}\rangle
N_{n',m}(E_3,E_2,\omega+\omega',t)\}\hphantom{ddd}& \label{crmf3}
\end{eqnarray}
The causal commutation relations of the master field are obtained replacing in (\ref{crmf1})--(\ref{crmf3}) the
factor $ \dl(t'-t) $ by $ \dl_+(t'-t) $, where the causal $\dl$-function $\dl_+(t'-t)$ is defined in
Ref.~\refcite{ALV}, section (8.4); $ 2\pi\dl(E_1-E_2-\omega) $ by $ (i(E_1-E_2-\om-i0))^{-1} $ and
$2\pi\dl(E_3-E_1\pm\omega)$ by $ (i(E_3-E_1\pm\om-i0))^{-1} $.
\end{theorem}

\subsection{Fock Representation of the Master Field}\label{sect3.2}
The next step is to realize the algebra of the master field
by operators acting in a Hilbert space. We realize it in the Fock space, which is constructed as follows.

Let $K$ be a vector space of finite rank operators acting on the one-particle Hilbert space ${\mathcal H}_1$ with
the property that for any $\omega\in\Omega$; $X,Y\in K$
\begin{eqnarray*}
\langle X,Y\rangle_\omega :&=&\int dt\Tr\left(e^{-\beta H_1}X^*S_tYS^*_t\right)e^{-i\omega t}
\equiv \int dt\Tr\left(e^{-\beta H_1}X^*Y_t\right)e^{-i\omega t}\\
&\equiv &2\pi\int dE\Tr\left(e^{-\beta H_1}X^*P_EYP_{E-\omega}\right)<\infty
\end{eqnarray*}
where $Y_t=S_tYS^*_t$ the free evolution of $Y$. It was shown in Ref.~\refcite{AL2} that the space $K$ is non
empty and $\langle \cdot,\cdot\rangle_\om$ defines a prescalar product on $K$. Let $K_\omega$ be the Hilbert space
with inner product $\langle\cdot,\cdot\rangle_\omega$ obtained as completion of the quotient of $K$ by the zero
$\langle\cdot,\cdot\rangle_\omega$-norm elements. Denote ${\mathcal
K}:=\bigoplus\limits_{\omega\in\Omega}K_\omega$.

Consider the Fock space
\begin{equation}\label{222}
  \Gamma(L^2(\mathbb R_+)\otimes{\mathcal K})\equiv
  \Gamma(\bigotimes\limits_{\om\in\Omega}L^2({\mathbb R}_+,K_\om))\equiv
  \bigotimes\limits_{\om\in\Omega}\Gamma(L^2({\mathbb R}_+,K_\om))
\end{equation}
where the last infinite tensor product is referred to the vacuum vectors. Let $b^+_{t,\omega}(X), b_{t,\omega}(X)
$ be the white noise creation and annihilation operators in this Fock space. These operators satisfy the
commutation relations
$$
 [b_{t,\omega}(X),b_{t',\omega'}^+(Y)]=\dl(t'-t)\dl_{\omega,\omega'}\langle X,Y\rangle_\omega.
$$
Each white noise operator $b_{t,\omega}(\cdot)$ acts as usual annihilation operator in $\Gamma(L^2(\mathbb R_+,
K_\omega))$ and as identity operator in other subspaces.

The representation of the algebra~(\ref{crmf1})--(\ref{crmf3}) can be constructed in the Fock space~(\ref{222}) by
the identification
$$
 B_{n,m}(E_1,E_2,\omega,t)=b_{t,\omega}(|P_{E_1}g_{n}\rangle\langle P_{E_2}g_{m}|)
$$
The number operator is defined as
\begin{eqnarray*}
&N_{n,m}(E_1,E_2,\omega,t)=&\\
&\sum\limits_{\ep=0,1}\sum\limits_{\omega_1\in\Omega} \mu_{\ep}(E_1-\omega_1)b^+_{t,\omega_1}(|g_{n}\rangle\langle
P_{E_1-\omega_1}g_{\ep}|) b_{t,\omega_1-\omega}(|P_{E_2}g_{m}\rangle\langle P_{E_1-\omega_1}g_{\ep}|)&
\end{eqnarray*}
with $\mu_\ep(E):=\langle g_\ep,P_Ee^{-\beta H_1}g_\ep\rangle^{-1}$. One easily checks that these operators
satisfy the commutation relations~(\ref{crmf1})--(\ref{crmf3}). Denote
\begin{eqnarray*}
& &N_{n,m}(\omega,t):=\int dE_1dE_2N_{n,m}(E_1,E_2,\omega,t)\\
& &B_{n,m}(E,\omega,t):=\int dE'B_{n,m}(E',E,\omega,t)
\end{eqnarray*}
The limiting white noise Hamiltonian acts in ${\mathcal H}_{\rm S}\otimes\Gamma(L^2(\mathbb R_+)\otimes{\mathcal
K})$ as
$$
 H(t)=\sum\limits_{\omega\in B}D_\omega\otimes\Bigl\{
 N_{0,1}(\omega,t)+\int dE\Bigl[B_{1,0}(E,-\omega,t)+
 B_{0,1}^+(E,\omega,t)\Bigr]\Bigr\}+h.c.
$$

\subsection{Quantum White Noise, Langevin and Boltzmann
Equations}\label{ssQWN} The white noise Schr\"odinger
equation for the evolution operator in the low density
limit is
\begin{equation}\label{equ}
\partial_tU_t=-iH(t)U_t
\end{equation}
Following the general theory of white noise equations, in order to give a precise meaning to this equation we will
put it in the causally normally ordered form, in which all annihilators are put on the right hand side of the
evolution operator and all creators are on the left hand side. This procedure gives a normally ordered quantum
white noise equation, which is equivalent to a quantum stochastic differential equation.

For any $n,m\in\{0,1\}$, $\om\in\Omega$, $E\in\mathbb R_+$ let $R^{n,m}_{\om,\om'}(E)$ be operators in ${\mathcal
H}_{\rm S}$ which are explicitly defined in Sec.~7, Ref.~\refcite{APV2}. The following theorem was proved in
Ref.~\refcite{APV2}.
\begin{theorem}
The normally ordered form of equation~(\ref{equ}) is
\begin{eqnarray}
& \partial_tU_t=\sum\limits_{n,m=0,1}\int dE\bigl[\, \sum\limits_{\om,\om'}
  R^{n,m}_{\om,\om'}(E)\sum\limits_{\ep=0,1}\mu_\ep(E)B^+_{n,\ep}(E,\om,t)
   U_tB_{m,\ep}(E,\omega',t)&\nonumber\\
& +\sum\limits_\om\Bigl(R^{n,m}_{\om,0}(E)B^+_{n,m}(E,\om,t)U_t+
  R^{m,n}_{0,\om}(E)U_tB_{n,m}(E,\om,t)\Bigr)&\nonumber\\
& +R^{n,m}_{0,0}(E)\langle g_{n},P_Ee^{-\beta H_1}g_{m}\rangle U_t\bigr]&\label{normordeq}
\end{eqnarray}
\end{theorem}

The normally ordered equation~(\ref{normordeq}) can be applied to derivation of the quantum Langevin equation for
test particle's observables. Let $X$ be any observable of the test particle. The Langevin equation is the equation
satisfied by the stochastic flow $j_t$ defined by $j_t(X)\equiv X_t :=U^+_tXU_t$.
\begin{theorem}
The quantity $X_t$ satisfies the quantum Langevin equation:
\begin{eqnarray}
& \dot X_t=\sum\limits_{n,m=0,1}\int dE\Bigl[\, \sum\limits_{\om_1,\om_2}
 \sum\limits_\ep\mu_\ep(E)B^+_{n,\ep}(E,\om_1,t)
 U^*_t\Theta^{n,m}_{\om_1,\om_2}(X)U_tB_{m,\ep}(E,\om_2,t)&\nonumber\\
& +\sum\limits_\om\Bigl(B^+_{n,m}(E,\om,t)U^*_t\Theta^{n,m}_{\om,0}(X)U_t+
 U^*_t\Theta^{m,n}_{0,\om}(X)U_tB_{n,m}(E,\om,t)\Bigr)&\nonumber\\
& +\langle g_{n},P_Ee^{-\beta H_1}g_{m}\rangle U^*_t\Theta^{n,m}_{0,0}(X)U_t\Bigr]&\label{Langevin}
\end{eqnarray}
where the structure maps are $\Theta^{n,m}_{\om_1,\om_2}(X):=XR^{n,m}_{\om_1,\om_2}(E)+ R^{+
m,n}_{\om_2,\om_1}(E)X+2\sum\limits_{\ep,\om}{\rm Re}\gamma_\ep(E+\om)
R^{+\ep,n}_{\om,\om_1}(E)XR^{\ep,m}_{\om,\om_2}(E)$.
\end{theorem}

Equation~(\ref{Langevin}) can be written in terms of the stochastic differentials:
\begin{eqnarray}
& dj_t(X)=j_t\circ\sum\limits_{n,m}\int dE\Bigl[\, \sum\limits_{\om_1,\om_2}
 \Theta^{n,m}_{\om_1,\om_2}(X)dN_t(Z^{n,m}_{\om_1,\om_2}(E))&\nonumber\\
& +\sum\limits_\om\Bigl(\Theta^{n,m}_{\om,0}(X)dB^+_t((|g_{n}\rangle\langle P_Eg_{m}|)_\om)+
 \Theta^{m,n}_{0,\om}(X)dB_t((|g_{n}\rangle\langle P_Eg_{m}|)_\om)\Bigr)\Bigr]&\nonumber\\
& +j_t\circ{\mathcal L}(X)dt&\nonumber
\end{eqnarray}
where $Z^{n,m}_{\om_1,\om_2}(E)$ is a certain operator in $\mathcal K$, which is explicitly defined in
Ref.~\refcite{APV2}; ${\mathcal L}$ is a quantum Markovian generator, which has the form of a generator of a
quantum dynamical semigroup\cite{gksl}:
\begin{equation*}
{\mathcal L}(X)=\Psi(X)-\frac{1}{2}\{\Psi(1),X\}+i[H_{\rm eff},X]
\end{equation*}
Here $\Psi(X)=2\pi\sum\limits_{\ep,\ep',\om}\int dE\langle
g_\ep,P_Ee^{-\beta H_1}g_\ep\rangle\langle
g_{\ep'}P_{E+\om}g_{\ep'}\rangle R^{+
\ep',\ep}_{\om,0}(E)XR^{\ep',\ep}_{\om,0}(E)$ is a
completely positive map and the effective Hamiltonian
$H_{\rm eff}:=\sum_{\ep}\int dE\langle g_\ep,P_Ee^{-\beta
H_1}g_\ep\rangle \times(R^{+
\ep,\ep}_{0,0}(E)-R^{\ep,\ep}_{0,0}(E))/2i$  is
selfadjoint.

The following theorem is a direct consequence of the
quantum Langevin equation~(\ref{Langevin}) of
Ref.~\refcite{APV2}.
\begin{theorem}\label{thb}
The reduced density matrix satisfies the quantum linear Boltzmann equation
\begin{equation}\label{eq1}
\frac{d\rho(t)}{dt}={\mathcal L}_*(\rho(t))
\end{equation}
where the generator ${\mathcal L}_*$ is the dual to the generator ${\mathcal L}$.
\end{theorem}

The generator ${\mathcal L}_*$ is the sum of its
dissipative and Hamiltonian parts, ${\mathcal
L}_*(\rho)={\mathcal L}_{\rm diss}(\rho)-i[H_{\rm
eff},\rho]$. The explicit form can be obtained by direct
calculations using the expression above for the generator
$\cal L$ and Eq.~(8.6) of Ref~\cite{APV2} for the $T$
operator. Finally it has the following form. Let $T$ be
the one-particle T-operator for the scattering of the test
particle and one particle of the gas and
$T_{n,n'}(k,k')=\langle n,k|T|n',k'\rangle$ be its generic
matrix element, where $|n\rangle$ is an eigenvector of
$H_S$ with eigenvalue $\ep_n$ and $k$ the momentum of one
particle of the gas. Denote
\[
T_\om(k,k'):=\sum\limits_{m,n:\, \ep_m-\ep_n=\omega}T_{m,n}(k,k')|m\rangle\langle n|
\]
Density of particles of the gas is determined by the function $L(k)$ in~(\ref{state}). For the Gibbs state
$L(k)=e^{-\beta\om(k)}$. Other forms of the density correspond to non-equilibrium states of the gas and can be
controlled, for example, by filtering. In these notations the dissipative part of the generator is
\begin{eqnarray}
&{\mathcal L}_{\rm diss}(\rho)=2\pi\sum\limits_{\om\in B}\int dkdk'\dl(\om(k')-\om(k)+\om)L(k)&\nonumber\\
&\times\Bigl[T_\om(k',k)\rho
T^+_\om(k',k)-\frac{1}{2}\Bigl(T^+_\om(k',k)T_\om(k',k)\rho+
\rho T^+_\om(k',k)T_\om(k',k)\Bigr)\Bigr]&\label{eq2}
\end{eqnarray}
In the case the gas is in equilibrium this generator coincides with the generator of the quantum linear Boltzmann
equation obtained in Ref.~\refcite{dumcke}.

\section{White Noise Approach without Fock-antiFock Representation}\label{sect4}
The approach to derivation of the quantum white noise equations directly in terms of the correlation functions,
without use of the Fock-antiFock representation, was developed in Ref.~\refcite{p}. In this approach one
introduces the notion of causal state and causal time-energy quantum white noise and proves the convergence of
chronological correlation functions of operators
\begin{equation}\label{2}
 N_{f,g,\xi}(t)=\frac{1}{\xi}A^+(S_{t/\xi}f)
 A(S_{t/\xi}g)
\end{equation}
acting in $\Gamma({\mathcal H}_1)$ to correlation functions of the time-energy quantum white noise
(Theorem~\ref{mastfi2}). This time-energy quantum white noise is a family of creation and annihilation operators,
with commutator proportional to $\dl$-function of time and energy [see~(\ref{ccrB0})]. These operators act in a
Fock space which, in difference with~(\ref{222}), does not depend on the initial state of the gas
$\varphi_{L,\xi}$.

Suppose for simplicity that $D(t)=D$. Then the evolution operator $U(t/\xi)$  after the time rescaling $ t\to
t/\xi $ satisfies the equation
\begin{equation}\label{eqev}
 \frac{dU(t/\xi)}{d t}=-i(D\otimes N_{g_0,g_1,\xi}(t)+D^+\otimes
 N_{g_1,g_0,\xi}(t))U(t/\xi)
\end{equation}
Theorem~\ref{mastfi2} and the causal commutation relations~(\ref{ccrB}) are used to show that the limit as $\xi\to
0$ of the rescaled evolution operator satisfies the causally normally ordered equation~(\ref{normordeq1}). That
equation is equivalent to the quantum stochastic equation~(\ref{qsde1}) which can be written in Hilbert module
notations as~(\ref{qsde2}) and then in terms of the one-particle $S$-matrix as~(\ref{qsde3}).

As it was stated in the Introduction, in this approach the algebra of the time-energy quantum white
noise~(\ref{ccrB0}), the quantum Ito table~(\ref{ito}) and the quantum stochastic equation for the limiting
evolution~(\ref{qsde3}) do not depend on the initial state of the gas. This is different from the approach of
Sect.~\ref{sect3}, where the commutation relations~(\ref{crmf1}) for the master field, the Hilbert space
representation (subsection~\ref{sect3.2}) and hence the limiting equation~(\ref{normordeq}) depend on the initial
state (through the factor $e^{-\beta H_1}$). Instead, the dependence on the initial state of the gas now is
contained in the limiting state $\varphi_L$ (in the equilibrium $L=e^{-\beta H_1}$) (in the approach of
Sect.~\ref{sect3} the state of the master field is the vacuum state). Considering the limiting state $\varphi_L$
as the conditional expectation [with the property~(\ref{st2})], one can derive the quantum master equation for the
reduced dynamics of the test particle which coincides, when restricted to the same model, with the analogous
equation following from~(\ref{Langevin}).

\subsection{Causal Time-Energy Quantum White
Noise}\label{ssCWN} Define the Hilbert space ${\mathcal
X}_{{\mathcal H}_1,H_1}$ as the completion of the quotient
of the set
\[
\left\{F: \mathbb R_+\to{\mathcal H}_1\,\,\,{\rm s.t.}\,\,\, ||F||^2:=2\pi\int\rmd E\langle
F(E),P_EF(E)\rangle<\infty\right\}
\]
with respect to the zero-norm elements. The inner product in ${\mathcal X}_{{\mathcal H}_1,H_1}$ is $\langle
F,G\rangle=2\pi\int\rmd E\langle F(E),P_EG(E)\rangle$. Let $B^+_f(E,t),\, B_g(E',t')$ be the creation and
annihilation operators acting in the symmetric Fock space $\Gamma(L^2(\mathbb R_+,{\mathcal X}_{{\mathcal
H}_1,H_1}))$ over the Hilbert space $L^2(\mathbb R_+,{\mathcal X}_{{\mathcal H}_1,H_1})$ of square integrable
functions $f: \mathbb R_+\to{\mathcal X}_{{\mathcal H}_1,H_1}$. These operators (operator-valued distributions)
satisfy the canonical commutation relations
\begin{equation}\label{ccrB0}
[B_g(E,t),\, B^+_f(E',t')]=2\pi\dl(t'-t)\dl(E'-E)\langle g,P_Ef\rangle
\end{equation}
and causal commutation relations
\begin{equation}\label{ccrB}
[B_g(E,t),\, B^+_f(E',t')]=\dl_+(t'-t)\dl(E'-E)\gamma_{g,f}(E)
\end{equation}
where $\dl_+(t'-t)$ is the causal $\dl$-function and $\gamma_{g,f}(E)=\int_{-\infty}^0 dt\langle g,S_tf\rangle
e^{-itE}$. The meaning of two different commutators~(\ref{ccrB0}) and~(\ref{ccrB}) for the same operators is
explained in Ref.~\refcite{ALV}, Sect.~7. These operators are called {\it time-energy quantum white noise} due to
the presence of $\dl(t'-t)\dl(E'-E)$ in~(\ref{ccrB0}).

Define the {\it white noise number operators} as
\begin{equation}\label{defNfg}
N_{f,g}(t)=\int dEB^+_f(E,t)B_g(E,t)
\end{equation}

For any positive bounded operator $L$ in ${\mathcal H}_1$ define the {\it causal gauge-invariant mean-zero
Gaussian state} $\varphi_L$ by the properties (\ref{prop1})--(\ref{state1}):
\begin{equation}\label{prop1}{\rm for\,\, n=2k}\qquad
\varphi_L(B^{\epsilon_1}_1\dots B^{\epsilon_n}_n)= \sum\varphi_L(B^{\epsilon_{i_1}}_{i_1}B^{\epsilon_{j_1}}_{j_1})
\dots\varphi_L(B^{\epsilon_{i_k}}_{i_k}B^{\epsilon_{j_k}}_{j_k})
\end{equation}
where the sum is taken over all permutations of the set $(1,\dots,2k)$ such that $i_\alpha< j_\alpha$,
$\alpha=1,\dots,k$ and $i_1<i_2<\dots<i_k$; $B^{\epsilon_m}_m:=B^{\epsilon_m}_{f_m}(E_m,t_m)$ for $m=1,\dots,n$
are time-energy quantum white noise operators with causal commutation relations~(\ref{ccrB}), and $\epsilon_m$
means either creation or annihilation operator;
\begin{equation}{\rm for\,\, n=2k+1}\qquad
\varphi_L(B^{\epsilon_1}_1\dots B^{\epsilon_n}_n)=0
\end{equation}
\begin{equation}
\varphi_L(B_f(E,t)B_g(E',t'))= \varphi_L(B^+_f(E,t)B^+_g(E',t'))=0
\end{equation}
\begin{equation}\label{state1}
\varphi_L(B^+_f(E,t)B_g(E,t'))=\chi_{[0,t]}(t')\langle g,P_ELf\rangle
\end{equation}
The "state" $\varphi_L$ does not satisfy the positivity condition. This is a well-known situation for the weak
coupling limit (see Ref.~\refcite{ALV}) and is due to the fact that we work with time-ordered, or causal
correlators. Therefore it is natural to call such "states" causal states.

\begin{definition}
{\it Causal time-energy quantum white noise} is a pair $(B^\pm_f(E,t),\varphi_L)$, where $B^\pm_f(E,t)$ satisfy
the causal commutation relations~(\ref{ccrB}) and $\varphi_L$ is a causal gauge-invariant mean-zero Gaussian
state.
\end{definition}
\begin{theorem}\label{mastfi2}
For any $ n\in\mathbb N$ in the sense of distributions over simplex $t_1\ge t_2\ge\dots \ge t_n\ge0$ one has the
limit
\[
\lim\limits_{\xi\to 0}\varphi_{L,\xi}(N_{f_1,g_1,\xi}(t_1)\dots
N_{f_n,g_n,\xi}(t_n))=\varphi_L(N_{f_1,g_1}(t_1)\dots N_{f_n,g_n}(t_n))
\]
\end{theorem}
This theorem was proved in Ref.~\refcite{p}.
\begin{remark}
This convergence is called convergence in the sense of time-ordered correlators. The fact that we use the
distributions over simplex is motivated by iterated series~(\ref{eqU111}) for the evolution operator.
\end{remark}
\begin{remark} The proof is based on the fact that for any $n\in\mathbb N$ only one connected diagram survives
in the limit. This can be interpreted as emergence of a
new statistics (different from Bose) in the low density
limit\footnote{In Ref.~\cite{p2} the statistics which
appears in the low density limit was found and a
connection with Voiculescu free independence theory was
established.}. For a discussion of new statistic arising
in the weak coupling limit see Ref.~\refcite{ALV}.
\end{remark}

The following theorem is important for investigation of the limiting white noise equation for the evolution
operator.
\begin{theorem}\label{limitstate} The limit state $\varphi_L$ has the following
factorization property: $\forall n\in\mathbb N$
\begin{eqnarray*}
&\varphi_L&(B^+_f(E,t)N_{f_1,g_1}(t_1)\dots
N_{f_n,g_n}(t_n)B_g(E,t))\\
&=& \varphi_L(B^+_f(E,t)B_g(E,t))\varphi_L(N_{f_1,g_1}(t_1)\dots N_{f_n,g_n}(t_n))
\end{eqnarray*}
where the equality is understood in the sense of distributions over simplex $t\ge t_1\ge t_2\ge\dots \ge t_n\ge0$.
\end{theorem}

Theorem~\ref{mastfi2} allows us to calculate the partial expectation of the evolution operator and Heisenberg
evolution of any system observable in the low density limit. In fact, partial expectation of the $n$-th term of
the iterated series~(\ref{eqU111}) (or equivalent series for Heisenberg evolution of a system observable) after
time rescaling $t\to t/\xi$ includes the quantity
\[
\int\limits_0^t\rmd t_1\dots\int\limits_0^{t_{n-1}}\rmd t_n \varphi_{L,\xi}(N_{f_1,g_1,\xi}(t_1)\dots
N_{f_n,g_n,\xi}(t_n))
\]
(where $f_\alpha,\, g_\alpha$ are equal to $g_0$ or $g_1$). The limit as $\xi\to 0$ of this quantity can be
calculated using Theorem~\ref{mastfi2}. For example, the contribution of the connected diagram is equal to
\begin{eqnarray}
\int\limits_0^t\rmd t_1\int\limits_0^{t_1}\rmd t_2\dl_+(t_2-t_1)\int\limits_0^{t_2}\rmd t_3\dl_+(t_3-t_2)\dots
\int\limits_0^{t_{n-1}}\rmd t_n\dl_+(t_n-t_{n-1})\nonumber\\
\times\int\rmd E
\langle g_n,P_ELf_1\rangle\gamma_{g_1,f_2}(E)\dots\gamma_{g_{n-1},f_n}(E)\nonumber\\
=t\int\rmd E \langle g_n,P_ELf_1\rangle\gamma_{g_1,f_2}(E)\dots\gamma_{g_{n-1},f_n}(E)\nonumber
\end{eqnarray}
Similarly one can calculate the contribution of nonconnected diagrams (they give terms proportional to higher
powers of $t$). Summation over all orders of the iterated series gives the reduced dynamics of the test particle.
An advantage of the white noise approach is that it allows to get the limiting dynamics in a nonperturbative way,
without direct summation of the iterated series. This procedure includes derivation of the causally normally
ordered white noise equation for the limiting evolution operator. After that the reduced dynamics of the test
particle can be easily found.

\subsection{The White Noise and Quantum Stochastic
Equations}\label{ssWNQSE} The limiting evolution operator
satisfies the white noise Schr\"odinger equation
\begin{equation}\label{equ1}
\frac{\rmd U_t}{\rmd t}=-\rmi (D\otimes N_{g_0,g_1}(t)+D^+\otimes N_{g_1,g_0}(t))U_t,\qquad U_0=1
\end{equation}
The next step is to put this equation to the causally normally ordered form, i.e., to put all annihilation
operators, appearing in $N_{f,g}(t)$, on the right side of the evolution operator and all creation operators on
the left side.

Assume that for each $ E\in {\mathbb R}$ the following inverse operators exist
\begin{eqnarray*}
 T_0(E):=\Bigl(1+\ga_{g_0,g_1}(E)D^+-\ga_{g_1,g_0}(E)D+(\ga_{g_0,g_0}\ga_{g_1,g_1}-
 \ga_{g_1,g_0}\ga_{g_0,g_1})(E)DD^+\Bigr)^{-1}\\
 T_1(E):=\Bigl(1+\ga_{g_0,g_1}(E)D^+-\ga_{g_1,g_0}(E)D+
 (\ga_{g_0,g_0}\ga_{g_1,g_1}-\ga_{g_1,g_0}\ga_{g_0,g_1})(E)D^+D\Bigr)^{-1}
\end{eqnarray*}
Denote
\begin{eqnarray}
R_{0,0}(E)&:=&\ga_{g_1,g_1}(E)DT_1(E)D^+,\quad R_{0,1}(E):=-DT_1(E)(1+\ga_{g_0,g_1}(E)D^+)\nonumber\\
R_{1,1}(E)&:=&\ga_{g_0,g_0}(E)D^+T_0(E)D,\quad R_{1,0}(E):=D^+T_0(E)(1-\ga_{g_1,g_0}(E)D)\nonumber
\end{eqnarray}

\begin{theorem}
The causally normally ordered form of equation~(\ref{equ1}) is
\begin{equation}\label{normordeq1}
\frac{\rmd U_t}{\rmd t}=-\sum\limits_{n,m=0,1}\int\rmd ER_{m,n}(E)B^+_{g_m}(E,t)U_tB_{g_n}(E,t)
\end{equation}
\end{theorem}
This theorem was proved in Ref.~\refcite{p}.
\begin{remark}
An immediate consequence of Theorem~\ref{limitstate} is the following factorization property of the limiting state
$\varphi_L$:
\[
\varphi_L(B^+_f(E,t)U_tB_g(E,t))= \varphi_L(B^+_f(E,t)B_g(E,t))\varphi_L(U_t)
\]
This property of the state $\varphi_L$ similar to the factorization property of the state determined by a coherent
vector $\Psi, \|\Psi\|=1$:
\[
(\Psi,B^+_f(E,t)U_tB_g(E,t)\Psi)= (\Psi,B^+_f(E,t)B_g(E,t)\Psi)(\Psi,U_t\Psi)
\]
which is usually used to define quantum stochastic differential equations.
\end{remark}

Normally ordered white noise equation~(\ref{normordeq1})
equivalent through identification
\[
B^+_m(E,t)U_tB_n(E,t)\rmd t=2\pi\rmd
N_t(|P_Eg_m\rangle\langle P_Eg_n|)U_t
\]
to the quantum stochastic differential equation
\begin{equation}\label{qsde1}
\rmd U_t=-2\pi\sum\limits_{n,m=0,1}\int\rmd ER_{m,n}(E)\rmd N_t(|P_Eg_m\rangle\langle P_Eg_n|)U_t
\end{equation}
where $N_t$ is the quantum number process in $\Gamma(L^2(\mathbb R_+)\otimes{\mathcal H}_1)$. The stochastic
differential $\rmd N_t$ satisfies the quantum Ito table
\begin{equation}\label{ito}
\rmd N_t(X)\rmd N_t(Y)=\rmd N_t(XY)
\end{equation}
where $X,Y$ are operators in ${\mathcal H}_1$. The limiting state $\varphi_L$ has the property
\begin{equation}\label{st2}
\varphi_L(2\pi\rmd {\rm N}_t(|P_Ef\rangle\langle P_Eg|))=\langle g,P_ELf\rangle\rmd t
\end{equation}

Equation~(\ref{qsde1}) can be written in Hilbert module notations as
\begin{equation}\label{qsde2}
\rmd U_t=\rmd N_t\Bigl(-2\pi\sum\limits_{n,m=0,1}\int\rmd ER_{m,n}(E)\otimes |P_Eg_m\rangle\langle
P_Eg_n|\Bigr)U_t
\end{equation}
The one-particle $S$-matrix for scattering of the test particle on one particle of the gas has the form
\begin{equation}\label{Smat}
S = 1 -2\pi\sum\limits_{n,m=0,1}\int\rmd ER_{m,n}(E)\otimes |P_Eg_m\rangle\langle P_Eg_n|
\end{equation}
This is a unitary operator: $S^+S=SS^+=1$. An immediate conclusion from~(\ref{qsde2}) and (\ref{Smat}) is the
following theorem which was proved in Ref.~\refcite{p} and is one of the main results of the paper.
\begin{theorem} The evolution operator in the low density limit satisfies
the quantum stochastic equation driven by the quantum number process with strength $S-1$:
\begin{equation}\label{qsde3}
\rmd U_t=\rmd N_t(S-1)U_t
\end{equation}
\end{theorem}

This equation can be applied to derivation of the quantum master equation for the reduced dynamics. Let $X\in
B({\mathcal H}_{\rm S})$ be an observable of the test particle. Its time evolution $X_t=U^+_tXU_t$ satisfies the
equation
\[
dX_t=dU^+_tXU_t+U^+_tXdU_t+dU^+_tXdU_t
\]
(we identify $X$ and $X\otimes 1$). Now, since
\[
dU^+_t=U^+_tdN_t(S^+-1)
\]
(we use the Hilbert module notations, hence $dN_t(S^+-1)$ does not commute with $U^+_t$) and using the quantum Ito
table~(\ref{ito}) one gets
\begin{eqnarray*}
dX_t&=&U^+_tdN_t(S^+-1)XU_t+U^+_tXdN_t(S-1)U_t\\
& &+U^+_tdN_t(S^+-1)XdN_t(S-1)U_t=U^+_tdN_t(\Theta(X))U_t
\end{eqnarray*}
where the map $\Theta: B({\mathcal H}_{\rm S})\to B({\mathcal H}_{\rm S}\otimes{\mathcal H}_1)$ has the form $
\Theta(X)=(S^+-1)X(S-1)+(S^+-1)X+X(S-1)\equiv S^+XS-X$. Simple computations, together with the explicit
form~(\ref{Smat}) for the $S$-matrix and the property $P_EP_{E'}=\dl(E-E')P_E$ of the projector, give the
expression
\[
\Theta(X)=2\pi\sum\limits_{n,m=0,1}\int dE\Theta^{n,m}_E(X)\otimes|P_Eg_m\rangle\langle P_Eg_n|
\]
where $\Theta^{n,m}_E(X)=2\pi\sum\limits_{n',m'}R^+_{n',m}(E)XR_{m',n}(E)\langle
g_{n'}P_Eg_{m'}\rangle-R^+_{n,m}(E)X-XR_{m,n}(E)$. Therefore
\[
dX_t=2\pi\sum\limits_{n,m=0,1}\int dE \rmd N_t(|P_Eg_m\rangle\langle P_Eg_n|)U^+_t\Theta^{n,m}_E(X)U_t
\]
The reduced dynamics $\overline{X}_t:=\varphi_L(X_t)\in B({\mathcal H}_{\rm S})$ is obtained by taking conditional
expectation of both sides of this equation in the state $\varphi_L$ and using~(\ref{st2}):
\[
\frac{d\bar X_t}{dt}=\sum\limits_{n,m=0,1}\int dE\langle g_n,P_ELg_m\rangle\overline{\Theta^{n,m}_E(X)}_t
\]
When restricted to the case of orthogonal form-factors
$g_0,g_1$, this master equation coincides with the one of
Ref.~\refcite{APV2}. The linear Boltzmann equation for the
reduced density matrix is the dual to this equation and
has the form of Eq.~(\ref{eq1}) with dissipative generator
of the form Eq.~(\ref{eq2}).

\section*{Appendix\footnote{This Appendix was added after publishing
the paper to briefly outline some new results in the
field.}} \addcontentsline{toc}{section}{Appendix}

This paper presents two versions of the white noise
approach to the investigation of the dynamics of a quantum
system interacting with a gaseous environment. The
approach allows to describe the dynamics of the total
system consisting of the particle and the environment in
the low density limit. The low density limit corresponds
to the kinetic regime when only pair collisions (i.e.,
collisions of the test particle at one time moment with
one particle of the gas) contribute to the dynamics such
that the probabilities of multi particle collisions are
negligible. Although the main result of the approach is
the quantum stochastic differential equation for the total
dynamics, quantum master equation for the reduced system
dynamics is derived as its simple consequence
[Eq.~(\ref{eq1}) with Lindblad-GKS generator given by
Eq.~(\ref{eq2})]. The derivation is performed {\it ab
initio} for general, including non-equilibrium, gases.

Section~2 of the present paper considers cases of discrete
and continuous spectrum of the test particle. The limiting
equations are explicitly derived here for the case of
discrete spectrum. This corresponds to confining the test
particle in a spatially finite region, which however can
be arbitrarily large. The master equation~(\ref{eq1})
describing the dynamics of a particle interacting with a
gas recently was applied to the problem of non-unitary
quantum control~\cite{PR}, where it was used to analyze
capabilities of controlling quantum systems by optimizing
with learning control algorithms the state of the
surrounding gas (i.e., by optimizing its distribution
function). Master equations for a quantum Brownian
particle in a free space (i.e., with translation invariant
dynamics) also attract attention of the researchers. In
this context I mention recent works by Bassano
Vacchini~\cite{BV}, Klaus Hornberger~\cite{KH}, and
Stephen Adler~\cite{Adler}, where also other relevant
references can be found. The two characteristic features
of the white noise approach are that it allows to derive
solvable equations for the dynamics of the total system
(the test particle and the environment) and that these
equations are not phenomenological, they are derived from
the exact microscopic dynamics.

\section*{Acknowledgments}
The first part of the paper (Sec.~\ref{sect3}) is based on
the joint work of the author with Professor L. Accardi and
Professor I.V. Volovich. The author is grateful to L.
Accardi for kind hospitality in the Centro Vito Volterra;
to L. Accardi, Y.G. Lu, and I.V. Volovich for many useful
and stimulating discussions. This work is partially
supported by a NATO-CNR Fellowship and Grant RFFI
02-01-01084.

\end{document}